\documentclass[a4paper,10pt]{article}

\usepackage[margin=25mm]{geometry}
\usepackage{times}
\usepackage{graphicx}
\usepackage{caption}
\usepackage{hyperref}
\usepackage{fancyhdr}
\usepackage{titlesec}
\titleformat{\section}{\bfseries\large\uppercase}{\thesection.}{1em}{}
\titleformat{\subsection}{\bfseries\normalsize}{\thesubsection}{1em}{}
\titleformat{\subsubsection}{\itshape\normalsize}{\thesubsubsection}{1em}{}

\renewenvironment{abstract}{
  \begin{center}
    {\large\bfseries Abstract \par} 
  \end{center}
}{\par\vspace{2em}}

\usepackage[sort&compress, numbers]{natbib}

\usepackage{amsmath,amsfonts,amssymb,amsthm} 
\usepackage{mathtools}
\usepackage{dsfont}
\usepackage{xcolor}
\usepackage{algpseudocode}
\usepackage{algorithm}
\usepackage{soul}
\usepackage{appendix}

\begin{document}

\title{\bfseries SELF-SUPERVISED CONFORMAL PREDICTION WITH EQUIVARIANT BOOTSTRAPPING FOR IMAGE UNCERTAINTY QUANTIFICATION}
\author{Henry J.~Aldridge$^{1}$, Tob{\'i}as I.~Liaudat$^{2}$, Marcelo Pereyra$^{3,4}$, Jason D.~McEwen$^{1,5}$ \\
{\small $^{1}$Mullard Space Science Laboratory, University College London, Dorking, UK, henry.aldridge.23@ucl.ac.uk}\\
{\small $^{2}$IRFU, CEA, Universit{\'e} Paris-Saclay, F-91191 Gif-sur-Yvette Cedex, France}\\
{\small $^{3}$Heriot-Watt University, Edinburgh, UK}\\
{\small $^{4}$Maxwell Institute for Mathematical Sciences, Edinburgh, UK}\\
{\small $^{5}$The Alan Turing Institute, London, UK}}
\date{}
\maketitle
\thispagestyle{fancy}

\begin{abstract}
  Inverse problems are ubiquitous in modern scientific studies and involve recovering an underlying signal from noisy observations often transformed by a measurement operator. These problems are frequently ill-posed, particularly in imaging, leading to multiple plausible solutions and considerable uncertainty in reconstructed images. In fields like the physical and biological sciences, accurate uncertainty quantification (UQ) is critical for trustworthy scientific analyses and confident diagnoses. Current UQ methods for imaging often fall short; they can be inaccurate, or require unavailable or difficult-to-acquire ground truth data for calibration, which can introduce hidden biases due to distribution shifts between calibration and observed data.
  We introduce a UQ approach that leverages equivariant bootstrapping to generate heuristic coverages by exploiting data symmetries. We then refine these coverages through a conformal prediction calibration step, while crucially employing a self-supervised approach to avoid the need for ground truth calibration data. We demonstrate this method with weak lensing mass-mapping, where we aim to reconstruct the convergence field from shear measurements of distant galaxies weakly-lensed by gravitational fields. Mass-mapping in particular benefits from the self-supervised approach, as simulating calibration data is expensive and relies on specific cosmological models that could introduce biases in downstream cosmological inference tasks.
\end{abstract}

\section{Introduction}

\paragraph{Inverse imaging problems}
Inverse imaging is the task of reconstructing an image ${x^{\star}} \in \mathbb{R}^{n}$ from a noisy observation ${y} \in \mathbb{R}^{m}$ that is a realisation of some stochastic process usually arising from a noisy acquisition process. We consider linear inverse imaging problems of the form
\begin{equation}
  {y} = \mathbf{A} {x^{\star}} + {\epsilon} ,
  \label{inverse_prob}
\end{equation}
where $\mathbf{A} \in \mathbb{R}^{m \times n}$ is a forward operator that models our measurement process and ${\epsilon} \sim \mathcal{N}(0,\Sigma)$ is additive Gaussian noise with covariance matrix $\Sigma$. Inverse imaging problems are associated with substantial reconstruction uncertainty due to ill-posed inverse mappings and noisy measurements, which can admit many plausible image reconstructions.

Reliable uncertainty quantification (UQ) in high-dimensional inverse problems is critical in imaging domains like medical diagnostics and astronomy, where uncertainty is tied to diagnoses or downstream scientific analysis. 
Existing UQ methods in inverse imaging problems often generate heuristic uncertainties that lack accuracy guarantees, or require ground truth data for calibration. We present a general method for performing image uncertainty quantification in a self-supervised manner and demonstrate it for the task of dark matter mass-mapping.

\paragraph{High-dimensional uncertainty quantification}
Despite strong progress with regards to recovering point estimates of ${x^{\star}}$, particularly with the advent of modern deep learning frameworks \cite{ongieDeepLearningTechniques2020}, estimating accurate and calibrated uncertainties in high dimensions remains a challenge. Existing approaches to UQ are typically formulated with a Bayesian statistical framework. These methods often exploit a physics-based likelihood term and employ different choices of prior models to regularise the ill-posed inverse problem. These priors can rely on simple assumptions like Gaussianity or sparsity in a given wavelet basis, or be data-driven and learned from large datasets of images. 
In either case, a recent study has shown that, in high dimensions, the probabilities of such models are poorly calibrated when confronted with a frequentist validation test \cite{thongBayesianImagingMethods2024}.

Conformal prediction \cite{angelopoulosGentleIntroductionConformal2022, angelopoulos2024theoretical} provides a statistically rigorous approach to calibrate uncertainties without assumptions on the error distribution and regardless of reconstruction method. Nevertheless, it relies on the availability of a set of exchangeable ground truth calibration data. 
In many applications, ground truth data are unavailable, expensive to simulate, or prone to distribution shifts relative to real observations. To ensure robustness in deployment, we must reduce reliance on ground-truth calibration data and move towards self-supervised methods.

A self-supervised approach to conformal prediction has recently been shown to provide accurate UQ consistent with its supervised counterpart in imaging problems with Gaussian noise statistics \cite{everinkSelfsupervisedConformalPrediction2025} and Poisson noise statistics \cite{tamoamougouSelfsupervisedConformalPrediction2025}. 
Given only a set of measurements and their reconstructions, the method utilises Stein's Unbiased Risk Estimator (SURE) \cite{steinEstimationMeanMultivariate1981} to provide an unbiased estimate of the mean squared error (MSE) between the unknown ground truth and the reconstructed images. Conformal prediction sets are constructed from quantile estimates of the MSE across the set of image reconstructions. Meanwhile, supervised conformalisation procedures have been applied to calibrate heuristic uncertainties, such as in \cite{angelopoulosImagetoImageRegressionDistributionFree2022,cherifUncertaintyQuantificationFast2024,letermeDistributionfreeUncertaintyQuantification2025,letermePlugandplayApproachFast2026}. Such approaches may generate per-pixel uncertainties and lead to more compact prediction sets.

Equivariant bootstrapping \cite{tachellaEquivariantBootstrappingUncertainty2024} provides a framework for generating per-image UQ, independent of the reconstruction method. The approach leverages symmetries in the imaging task to generate compact confidence regions in situations that are highly ill-posed due to large nullspaces of the forward operator. In CARB \cite{cherifUncertaintyQuantificationFast2024}, the equivariant bootstrapping approach was combined with a conformal prediction step to calibrate the resulting UQ. However, the method still relies on the availability of a calibration dataset, which may not be possible to obtain in a realistic setting, and no self-supervised approach has been explored.

\paragraph{Contributions}
Our approach aims to combine the method-agnostic uncertainty quantification of the equivariant bootstrapping with a self-supervised SURE-based conformal prediction step. We follow \cite{cherifUncertaintyQuantificationFast2024} in using Risk-Controlling Prediction Sets (RCPS) \cite{batesDistributionFreeRiskControllingPrediction2021, angelopoulosImagetoImageRegressionDistributionFree2022}, a conformal prediction approach that provides guaranteed coverage to a user-specified level of error, to calibrate the confidence regions provided by equivariant bootstrapping. Notably, the proposed UQ framework is fully self-supervised, requiring no ground truth calibration data and therefore avoiding distribution shifts. We demonstrate the method for weak gravitational lensing mass-mapping, a cosmological inverse imaging problem that aims to reconstruct the dark matter convergence field from noisy shear observations of distant galaxies. Existing UQ methods for mass-mapping often rely on simulated data for training or calibration, which may introduce hidden biases when exposed to real observations. Our self-supervised approach avoids this issue while providing statistically rigorous uncertainty quantification.

We continue this article by introducing the weak gravitational lensing mass-mapping inverse problem in Section \ref{sc:massmapping}. In Section \ref{sc:methodology} we present our proposed self-supervised UQ framework. In Section \ref{sc:results} we demonstrate the method for weak lensing mass-mapping and we conclude with final remarks in Section \ref{sc:conclusion}.

\subsection{The weak gravitational lensing mass-mapping inverse problem}
\label{sc:massmapping}

We demonstrate our proposed method with a cosmological application based on weak gravitational lensing, which probes the distribution of dark matter from galaxy observations. The photons emitted by these distant galaxies are weakly lensed by the gravitational potential of intervening matter (predominantly dark matter). By measuring the ellipticities of these galaxies, we obtain an estimate of the cosmic shear map, which constitutes the observation of the mass-mapping inverse problem. The convergence map, describing the extent of magnification of an object in that area of the sky, can then be inferred from the observed shear and forms a key component of modern cosmological analyses \cite{mandelbaumWeakLensingPrecision2018}. Large scale surveys such as the ongoing \textit{Euclid} space mission \cite{laureijsEuclidDefinitionStudy2011} and the Vera C. Rubin Observatory \cite{ivezicLSSTScienceDrivers2019} will provide vast amounts of data for unprecedented precision levels of cosmological analysis. It is therefore imperative that UQ for weak lensing mass-mapping is accurate to avoid biases in the estimation of cosmological parameters. Existing UQ approaches are often tied to specific sets of cosmological parameters by the training of data-driven prior models on simulated data \cite{remyProbabilisticMassmappingNeural2023,whitneyGenerativeModellingMassmapping2025} or by performing conformal prediction with a set of simulated calibration data \cite{letermeDistributionfreeUncertaintyQuantification2025,letermePlugandplayApproachFast2026}. These methods are therefore prone to exhibit hidden biases when exposed to distribution shifts in the observational data. Other mass-mapping methods with UQ do not rely on simulation-trained models, 
but they rely on simple priors like sparsity in a wavelet domain \cite{priceSparseBayesianMass2019} or Gaussian assumptions \cite{wienerExtrapolationInterpolationSmoothing1949}. These choices of priors are not complex enough to faithfully represent the underlying convergence field and can also provide uncalibrated probabilities.

In mass-mapping, we aim to recover the convergence map $\kappa \in \mathbb{R}^{m}$ from the observed shear map $\gamma \in \mathbb{C}^{m}$. The forward measurement process can be recast as a convolution with a lensing operator in the Fourier domain, along with additive Gaussian noise dependent on the standard deviation of galaxy ellipticities and number of galaxies observed per pixel. Following the formalism of (\ref{inverse_prob}), our inverse problem can be expressed as 
\begin{equation}
  {\gamma} = \mathbf{F}^{-1}\mathbf{DF}{\kappa} + {\epsilon} \qquad \text{where} \qquad \mathbf{D}=\frac{k_{x}^{2}-k_{y}^{2}+2i k_{x}k_{y}}{k_{x}^{2}+k_{y}^{2}}.
  \label{massmapping}
\end{equation}
Here, $\mathbf{D}$ is our Fourier-domain lensing operator for Cartesian co-ordinates $x, y$ parametrising the sky, $\mathbf{F}$ and $\mathbf{F}^{-1}$ are Fourier and inverse Fourier transforms, $k_{x}$ and $k_{y}$ are the Fourier frequencies and $\epsilon \sim \mathcal{CN}(0, \Sigma)$ with $\epsilon \in \mathbb{C}^{m}$, where $\Sigma$ is diagonal. The forward operator, $\mathbf{A}=\mathbf{F}^{-1}\mathbf{DF}$, is almost full-rank since only the DC Fourier mode is unobservable.
For more detail on weak gravitational lensing and mass-mapping, see \cite{bartelmannWeakGravitationalLensing2001,mandelbaumWeakLensingPrecision2018}. The classical approach to solving (\ref{massmapping}) is known as Kaiser-Squires \cite{kaiserMappingDarkMatter1993} and involves simply inverting the lensing operator and applying a smoothing operation. Kaiser-Squires leads to poor reconstructions due to amplification of noise in the inversion resulting in loss of small-scale structure when subsequently smoothed. Research into more advanced reconstruction methods is ongoing; modern approaches typically involve wavelet sparsity regularisation \cite{lanusseHighResolutionWeak2016,priceSparseBayesianMass2019,priceSparseBayesianMass2020,priceSparseBayesianMassmapping2020a,priceSparseBayesianMass2021a,starckWeaklensingMassReconstruction2021} and deep learning \cite{jeffreyDeepLearningDark2020,shirasakiNoiseReductionWeak2021,remyProbabilisticMassmappingNeural2023,whitneyGenerativeModellingMassmapping2025}. Our method-agnostic UQ framework provides calibrated uncertainty estimations and complements the ongoing research of powerful reconstruction-methods. For the sake of demonstrating our approach, we use the simple but widely adopted Kaiser-Squires method for reconstruction.

\section{Methodology}
\label{sc:methodology}

We propose a UQ framework for inverse imaging problems that generates statistically rigorous per-image measures of uncertainty via a self-supervised conformalisation step.

By considering a bootstrapping approach, in a method-agnostic framework we can generate compact heuristic per-image uncertainties without requiring training data. Straightforward parametric bootstrapping explores the variability of a reconstruction model for a given image reconstruction. Given some observation ${y}$, we generate an estimate ${\hat{x}(y)}$ using our reconstruction method of choice. Then, we generate a set of $N_B$ bootstrap observations $\{{\tilde{y}_i(\hat{x})}\}^{N_B}_{i=1}$ from our estimator by passing it through the stochastic process $y \sim P(\mathbf{A}x^{\star})$ described by (\ref{inverse_prob}), where $P(\cdot)$ represents the noise distribution. We pass the bootstrap observations through our chosen reconstruction method to generate $N_B$ bootstrap samples $\{{\tilde{x}(\tilde{y}_i)}\}^{N_B}_{i=1}$. We then compute a set of non-conformity scores $\{s({\tilde{x}}_i, {\hat{x}(y)})\}^{N_B}_{i=1}$ between each bootstrap sample and estimator, e.g.\ whole-image statistics such as the MSE or pixel-wise residuals. By considering the region for which the top $\alpha$-quantile has a score $s$ less than $q_\alpha$, we construct confidence regions $\mathcal{C}_\alpha({\hat{x}}) = \{x : s(x, \hat{x}(y)) < q_{\alpha} \}$ for $x^{\star}$. This serves as a measure of uncertainty for our reconstruction $\hat{x}(y)$.

However, parametric bootstrapping has poor performance in imaging settings due to the bias inherent in bootstrapping from a point estimate, particularly in ill-posed settings. Therefore, we adopt an approach tuned to imaging settings. Equivariant bootstrapping \cite{tachellaEquivariantBootstrappingUncertainty2024} adapts the parametric approach by leveraging symmetries in the imaging problem, leading to more performant imaging UQ when the problem is ill-posed. Given a set of signals $\mathcal{X}$, for which ${x}^{\star} \in \mathcal{X}$, equivariant bootstrapping exploits the set of invertible transformations $\mathcal{G}$ that leave $\mathcal{X}$ invariant, while typically the forward operator $\mathbf{A}$ or the reconstruction method $\hat{x}(y)$ are not $\mathcal{G}$-equivariant. Transformations $T_{g_i}$ are randomly sampled from $\mathcal{G}$ 
and applied to our estimator ${\hat{x}}(y)$ in the process of obtaining a realisation of a bootstrap observation ${\tilde{y}}_i \sim P(\mathbf{A}T_{g_i}{\hat{x}})$. Bootstrap samples are then recovered as ${\tilde{x}}_i = T^{-1}_{g_i}{\hat{x}}({\tilde{y}}_i)$. If $\mathbf{A}$ is not $\mathcal{G}$-equivariant, the transformed forward operators $\mathbf{A}T_{g_i}$ have overlapping but distinct nullspaces, allowing the bootstrap to probe uncertainty beyond what $\mathbf{A}$ alone can constrain. This requires careful selection of the transformations $\mathcal{G}$ for the problem at hand. We note that in our self-supervised implementation that follows, we measure error only in the subspace constrained by the forward operator, excluding components in the nullspace of $\mathbf{A}$ that are unobservable from the measurements. This is not in line with the original motivations behind equivariant bootstrapping to produce UQ inclusive of the nullspace of the forward operator. However, we argue that equivariant bootstrapping exposes the non-equivariance of the chosen reconstruction model under transformations $\mathcal{G}$, and thus generates variability in (\ref{score}) that better reflects uncertainty than parametric bootstrapping otherwise would.

To achieve statistically accurate coverages, we calibrate our heuristic confidence regions in a finite-sample setting using Risk-Controlling Prediction Sets (RCPS) \cite{batesDistributionFreeRiskControllingPrediction2021}, an advanced conformal prediction method, providing a framework with guaranteed coverage to a user-specified level of error. RCPS has proven successful in previous general imaging applications \cite{angelopoulosImagetoImageRegressionDistributionFree2022} and equivariant bootstrapping \cite{cherifUncertaintyQuantificationFast2024}. Given some target risk level $\alpha \in (0,1)$ and error rate $\delta \in (0,1)$, calibrated prediction sets contain the fraction $1-\alpha$ of possible predictions with probability $1-\delta$. This guarantee requires that calibration and test observations are exchangeable. In our self-supervised setting, where no ground truth is used, this reduces to the requirement that calibration and test observations are drawn from the same measurement distribution. RCPS redefines confidence regions $\mathcal{C}_{\alpha}({\hat{x}})$ obtained from equivariant bootstrapping as $\mathcal{C}_{\lambda;\alpha}({\hat{x}}) = \{x : s(x, {\hat{x}}) < \lambda q_{\alpha} \}$ where $\lambda \in \mathbb{R}_+$ is a scalar value adjusted from the RCPS calibration. RCPS seeks to find the smallest $\lambda$ that scales the prediction set $\mathcal{C}_{\lambda;\alpha}$ to meet its target ($\alpha, \delta$)-RCPS calibration, that is defined as follows
\begin{equation}
  \mathbb{P}(R(\mathcal{C}_{\lambda;\alpha}({\hat{x}})) \le \alpha) \ge 1 - \delta ,
  \label{RCPS_prob}
\end{equation}
where $R(\mathcal{C}_{\lambda;\alpha}({\hat{x}}))$ is the risk associated with the set $\mathcal{C}_{\lambda;\alpha}({\hat{x}})$. This risk is modelled as an expectation of some loss function $R(\mathcal{C}_{\lambda;\alpha}({\hat{x}})) = \mathbb{E}\left[L({x}^{\star}, \mathcal{C}_{\lambda;\alpha}({\hat{x}}))\right]$, where the loss $L({x}^{\star}, \mathcal{C}_{\lambda;\alpha}({\hat{x}}))$ is a monotonic measure of how wrong our prediction sets are. In the case of a whole-image non-conformity score $s({\cdot, \hat{x}})$, such as the MSE, our loss function can be a simple indicator function $L({x}^{\star}, \mathcal{C}_{\lambda;\alpha}({\hat{x}})) = \mathds{1}_{x^{\star} \notin \mathcal{C}_{\lambda;\alpha}}$ e.g.\ {does the true image lie outside the prediction set $\mathcal{C}_{\lambda;\alpha}({\hat{x}})$?} An upper confidence bound (UCB) of the risk, $\widehat{R}^{+}(\mathcal{C}_{\lambda;\alpha}({\hat{x}}))$, can be estimated from the empirical risk, $\widehat{R}(\mathcal{C}_{\lambda;\alpha}({\hat{x}}))$, computed from an empirical distribution of $N_C$ losses measured with a set of $N_C$ calibration data. Subsequently, $\lambda$ is tuned to meet the requirement of (\ref{RCPS_prob}) via a root-finding algorithm that searches for the smallest $\lambda$ satisfying the constraint. For a loss that takes binary values $\{0,1\}$, such as ours, the UCB can be precisely derived from an exact binomial result \cite[Theorem B.1]{batesDistributionFreeRiskControllingPrediction2021}, leading to the following UCB
\begin{equation}
  \widehat{R}^{+}(\mathcal{C}_{\lambda;\alpha}({\hat{x}}))=\mathrm{sup}\left\{R:P\left(\mathrm{Binom}(N_C,R)\leq\lceil N_C \widehat{R}(\mathcal{C}_{\lambda;\alpha}({\hat{x}}))\rceil\right)\geq\delta\right\},
  \label{UCB}
\end{equation}
where $\lceil \cdot \rceil$ denotes the ceiling function. See \cite{batesDistributionFreeRiskControllingPrediction2021} for further details on (\ref{UCB}) and UCBs for other loss functions.

By leveraging a generalised form of Stein's unbiased risk estimate (SURE) \cite{steinEstimationMeanMultivariate1981}, we can alleviate the requirement of ground truth calibration data for RCPS, yielding an approach for image uncertainty quantification that is fully self-supervised. We consider the non-conformity score
\begin{equation}
s_{\mathbf{M}}({x}^{\star}, {\hat{x}}({y})) = \frac{1}{m}\| {x}^{\star}- {\hat{x}}({y})\|^2_{\mathbf{M}} = \frac{1}{m}({x}^{\star}- {\hat{x}}({y}))^{\top} \mathbf{M} ({x}^{\star}- {\hat{x}}({y})),
\end{equation}
with $\mathbf{M}$ being a positive semi-definite matrix that allows us to define prediction sets that are more compact and informative for the inverse problem at hand. A natural choice is an approximation of the inverse-covariance of the error ${x}^{\star}- {\hat{x}}({y})$ \cite{johnstoneConformalUncertaintySets2021a}, such as $\mathbf{M} = \mathbf{A}^{\top} \mathbf{A}$, minimising score variability across images and thus yielding compact prediction sets. Consequently, the non-conformity score can be rewritten as
\begin{equation}
  s_{\mathbf{A}^{\top}\mathbf{A}}({x}^{\star}, {\hat{x}}({y})) = \frac{1}{m} \| \mathbf{A}{x}^{\star}- \mathbf{A}{\hat{x}}({y})\|^2_{2}.
  \label{score}
\end{equation}  
Given our observations are Gaussian distributed $y \sim \mathcal{N}(\mathbf{A}x^{\star}, \Sigma)$, we can use the SURE estimator to provide an unbiased estimate of the non-conformity score $s_{\mathbf{A}^{\top} \mathbf{A}}({x}^{\star}, {\hat{x}}({y}))$. The SURE expression is the following
\begin{equation}
  \text{SURE}(y)=\frac{1}{m}\|y-\mathbf{A}\hat{x}(y)\|_{2}^{2}-\text{tr}(\Sigma)+\frac{2}{m}\text{tr}\left(\Sigma\frac{\partial}{\partial y}\mathbf{A}\hat{x}(y)\right).
  \label{SURE}
\end{equation}
The traces in (\ref{SURE}) can be efficiently computed with the Skilling-Hutchinson trace estimator \cite{skillingEigenvaluesMegadimensionalMatrices1989,hutchinsonStochasticEstimatorTrace1989}, utilising automatic differentiation to compute the necessary gradients in the third term. Additionally, the covariance matrix $\Sigma$ reduces to a scalar factor $\sigma^2$ outside of the traces when $\epsilon \sim \mathcal{N}(0, \sigma^2\mathds{1}_m)$ in our inverse problem (\ref{inverse_prob}), and the trace on the right-hand side becomes a divergence. 

An alternative is to choose $\mathbf{M} = \mathbf{P} = \mathbf{A}^{\dagger}\mathbf{A}$ \cite{eldarGeneralizedSUREExponential2009, tachellaUNSURESelfsupervisedLearning2025}, i.e. the orthogonal projector onto the row space of $\mathbf{A}$, where $\mathbf{A}^{\dagger}$ denotes the Moore-Penrose pseudoinverse. This score measures error directly in the image-space, while (\ref{score}) in the measurement-space. However, for the purpose of conformal prediction whereby we wish to construct prediction sets that are informative and avoid over- and undercoverage, we opt for the score (\ref{score}) following \cite{everinkSelfsupervisedConformalPrediction2025}. We provide more detail on this alternative score and its SURE implementation in Appendix \ref{appendix}. As noted previously, both scores, approximated with their SURE counterparts, are blind to error in the nullspace of $\mathbf{A}$.

In the context of mass-mapping, shear observations are complex while the SURE formulation (\ref{SURE}) is applicable only to real signals. We thus consider $y = \gamma \in \mathbb{C}^{m} \mapsto \mathbb{R}^{2m}$, i.e.\ we consider a real field with twice as many dimensions, and $m\mapsto 2m$ in (\ref{SURE}) accordingly. Additionally, $\mathbf{A}^{\top}\mathbf{A} \approx \mathbf{I}$  (up to the unobservable DC mode), so the unnormalised measurement-space score $\|\mathbf{A}x^{\star}-\mathbf{A}\hat{x}(y)\|_{2}^{2}$ is approximately equivalent to the image-space error $\|x^{\star}-\hat{x}(y)\|^2_2$, and the prediction sets constructed from (\ref{score}) are approximately equivalent to those constructed from the image-space MSE. In the general setting, our method uses (\ref{score}) as the non-conformity score for equivariant bootstrapping and estimates its empirical distribution via SURE (\ref{SURE}) to enable self-supervised RCPS calibration. Our implementation conforms with the DeepInverse library \cite{tachella2025deepinverse} for easy implementation with other inverse imaging problems and we make our codebase available\footnote{\url{https://github.com/astro-informatics/sscb-imaging}}.

\section{Experimental Results}
\label{sc:results}

\begin{figure}
  \centering
  \includegraphics[width=\textwidth]{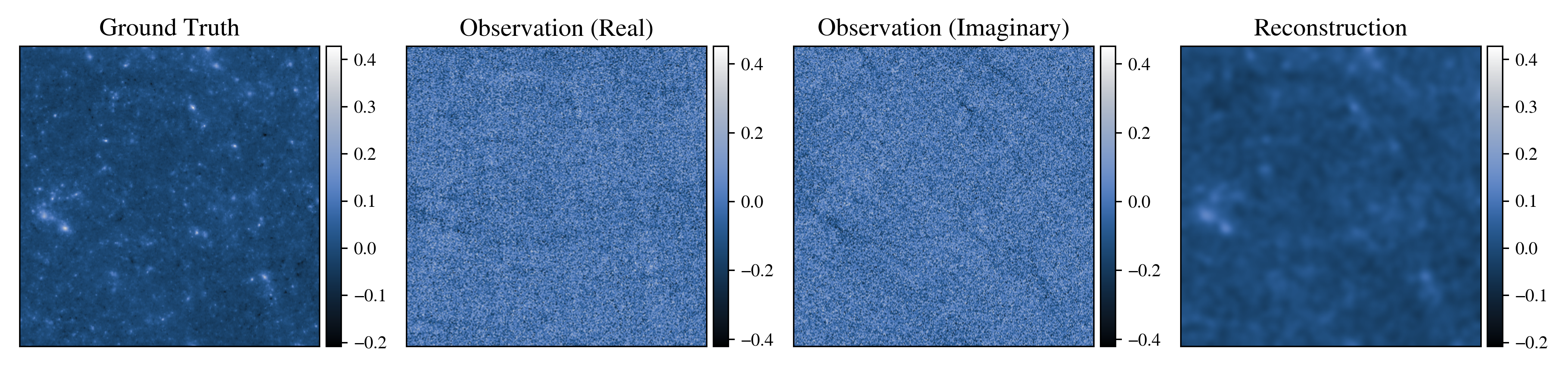}
  \caption{The weak lensing mass-mapping inverse imaging problem. Ground truth corresponds to simulated convergence maps. Noisy shear map observations are visualised in terms of real and imaginary components. The convergence map is reconstructed from the observed shear with Kaiser-Squires.}
  \label{massmapping_demo}
\end{figure}

To demonstrate the approach to weak lensing mass-mapping, we follow the methodology in \cite{whitneyGenerativeModellingMassmapping2025} to generate a set of $300\times300$ pixel mock COSMOS survey \cite{scovilleCOSMOSHubbleSpace2007} convergence maps based upon the $\kappa$TNG suite of convergence map simulations \cite{osatoKTNGEffectBaryonic2021}. We generate mock shear observations by applying the forward process in (\ref{massmapping}) to the simulated convergence maps. Rather than following \cite{whitneyGenerativeModellingMassmapping2025} in applying the realistic spatially varying noise, we simplify our setting and apply a constant Gaussian noise with a standard deviation averaged over the per-pixel standard deviations, and we do not apply the real COSMOS survey mask. Following \cite{whitneyGenerativeModellingMassmapping2025}, Kaiser-Squires image reconstructions are smoothed with a Gaussian kernel of $\sigma = 1 / 0.29$ pixels. Our mass-mapping inverse imaging problem is illustrated in Fig. \ref{massmapping_demo}.

We follow the implementation of equivariant bootstrapping in \cite{cherifUncertaintyQuantificationFast2024} and randomly sample 
compositions of transformations including vertical and horizontal flips, $90$ degree rotations, and cyclic image shifts of up to 2 pixels in both horizontal and vertical directions.
The approach additionally uses low and high frequency shelving filters where frequencies are attenuated to some fixed small fraction of their original value, maintaining the transformation invertibility necessary to recover bootstrap samples. 
We apply such filters, attenuating frequencies to $5\%$ of their original amplitude above and below frequency thresholds sampled from a Gaussian distribution tuned to our mass-mapping problem: we sample threshold frequencies centered around $200$ and $350$, with standard deviations of $50$. We include a constraint to prevent the low frequency threshold being larger than the high frequency threshold.
The geometric transformations leave the convergence field invariant, justified by the statistical isotropy and homogeneity of the field  in the flat-sky setting, while the forward operator is not equivariant under these transformations due to its directional dependence. The shelving filters, while relaxing the strict invariance requirement, complement the geometric transformations by probing scale-dependent reconstruction sensitivity and are empirically observed to improve the quality of the generated confidence regions.

\begin{figure}
  \centering
  \includegraphics[width=0.65\textwidth]{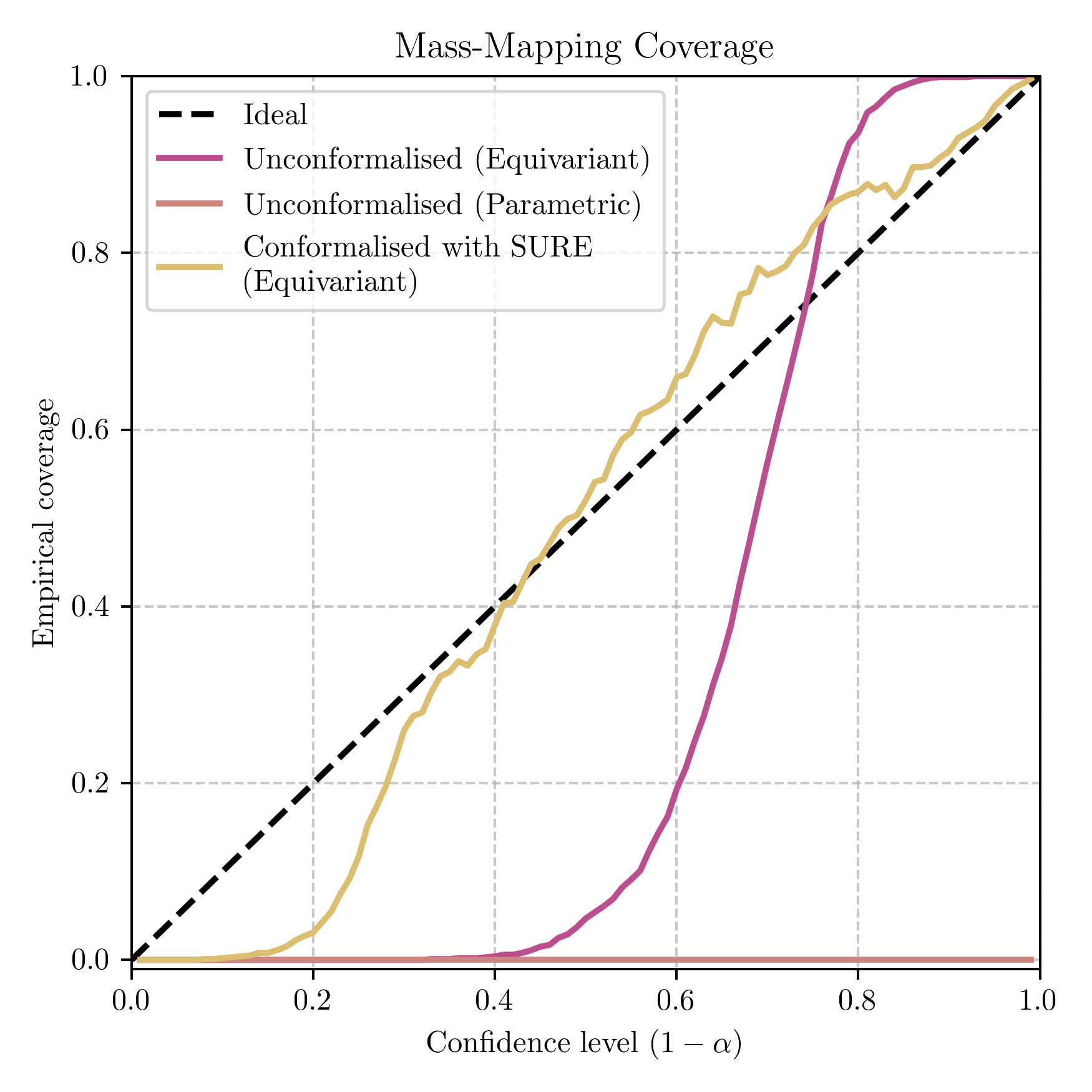}
  \caption{Coverage test performed with and without self-supervised conformalisation with SURE. Unconformalised corresponds to performing only equivariant or parametric bootstrapping.}
  \label{cov_test}
\end{figure}

We evaluate coverage at confidence levels with increments of 0.01, whereby the proportion of test samples lying within the predicted confidence regions are measured. We randomly select $1000$ mock shear observations to generate $1000$ convergence reconstructions. Our equivariant bootstrapping implementation generates $100$ bootstrap samples per image reconstruction to generate heuristic confidence regions with respect to the non-conformity score (\ref{score}), which are then conformalised with RCPS using SURE and error rate $\delta=0.1$. 
The scaling factors $\lambda$ generated by RCPS for a given confidence level are then tested on a separate test set of $1000$ mock observations. The results of the coverage test are shown in Fig. \ref{cov_test}. The conformalised regions closely track the ideal coverage diagonal. This is notable given that calibration was performed entirely without ground truth data. Deviation above and below the diagonal is permitted by our RCPS formulation (\ref{RCPS_prob}), and can be attributed to finite calibration and test set sizes, along with estimator noise inherent to SURE. In the unconformalised setting, equivariant bootstrapping outperforms parametric bootstrapping, which provides negligible coverage across all confidence levels, reflecting its inability to capture adequate reconstruction variability.

\section{Conclusion}
\label{sc:conclusion}

We present a general method for performing calibrated UQ in inverse imaging problems that is entirely self-supervised, alleviating the need to rely on ground truth data that may be simulated or difficult to acquire, and agnostic to the choice of reconstruction method. Therefore, our framework not only generates statistically accurate confidence regions but avoids risk of distribution shift between calibration data and observation data. This is achieved by utilising equivariant bootstrapping, which exploits data symmetries to generate heuristic uncertainties, thus avoiding any data-driven approach, and is then calibrated in a self-supervised manner with Stein's unbiased risk estimate. Our framework differs from \cite{everinkSelfsupervisedConformalPrediction2025} in that our bootstrapping approach generates bootstrap quantiles that vary per-image, providing more compact prediction sets. We demonstrate accurate coverage results for weak gravitational lensing mass-mapping, a cosmological inverse imaging problem used to constrain cosmological parameters. Existing UQ methods in this domain either rely on simulated data tied to specific cosmological models or parameters, or are uncalibrated, and therefore lack robustness when deployed.

Our approach has several limitations that should guide future work. The calibration step requires access to multiple observations that are exchangeable under the underlying distribution. In our demonstration we assumed access to such observations, whereas in real settings weak lensing mass-mapping will prefer the use of one complete observation of the sky to account for large-scale structures. While shear emulations can be generated from a single observation, such emulations are not necessarily exchangeable with independent samples from the true cosmological distribution, and the impact of this violation on the coverage guarantee warrants further investigation. Furthermore, the framework quantifies uncertainty at the image level (MSE), while existing supervised mass-mapping UQ approaches often provide per-pixel residual uncertainty estimates. More generally, our SURE formulation is blind to error in the nullspace of the forward operator. Therefore, as the rank-deficiency of the forward operator grows, there is greater possibility of underestimating the error in our reconstructions. This is not an issue for mass-mapping where our forward operator is almost full-rank, however in other inverse imaging problems this can become a more significant issue. A self-supervised framework for UQ in severely rank-deficient settings must address the insensitivity of the score to error in the nullspace. Additionally, the computational cost of generating many bootstrap samples favours fast reconstruction techniques, but conversely highlights the importance of continuing the development of efficient reconstruction algorithms. Beyond these limitations, the framework could be extended to support other noise models using generalized SURE formulations \cite{eldarGeneralizedSUREExponential2009}.

\begin{appendices}
\section{Image-space Score}
\label{appendix}
Choosing $\mathbf{M} = \mathbf{P} = \mathbf{A}^{\dagger}\mathbf{A}$, since $\mathbf{P}$ is an orthogonal projector ($\mathbf{P}^2 = \mathbf{P}$, $\mathbf{P}^{\top} = \mathbf{P}$), the non-conformity score can be rewritten as
\begin{equation}
  s_{\mathbf{P}}({x}^{\star}, {\hat{x}}({y})) = \frac{1}{m} \| \mathbf{P}{x}^{\star}- \mathbf{P}{\hat{x}}({y})\|^2_{2} = \frac{1}{m} \| \mathbf{A}^{\dagger}\mathbf{A}{x}^{\star}- \mathbf{A}^{\dagger}\mathbf{A}{\hat{x}}({y})\|^2_{2},
  \label{score_2}
\end{equation}
leading to the SURE expression
\begin{equation}
  \text{SURE}(y)= \frac{1}{m}\|\mathbf{A}^{\dagger}(\mathbf{A}\hat{x}(y)-y)\|^2_2 - \frac{1}{m}\text{tr}\left((\mathbf{A}^{\dagger})^{\top}\mathbf{A}^{\dagger}\Sigma\right) + \frac{2}{m}\text{tr}\left(\Sigma \frac{\partial}{\partial y} ((\mathbf{A}^{\dagger})^{\top} \mathbf{A}^{\dagger} \mathbf{A}\hat{x}(y))\right).
  \label{SURE_2}
\end{equation}
In the context of mass-mapping, our forward operator $\mathbf{A} = \mathbf{F}^{-1}\mathbf{DF}$ has pseudoinverse $\mathbf{A}^{\dagger} = \mathbf{A}^{\ast}$. Otherwise, if no closed-form solution exists, $\mathbf{A}^{\dagger}$ may be computed using singular vector decomposition, or its operation can be formulated as an optimisation problem. Additionally, the adjoint $(\mathbf{A}^{\dagger})^{\top}$ can be computed via automatic differentiation \footnote{As implemented in the DeepInverse library: \url{https://deepinv.github.io/deepinv/api/stubs/deepinv.physics.adjoint_function.html}}.
\end{appendices}

\bibliographystyle{IEEEtranN}   
\bibliography{bibliography/MaxEnt2025.bib}

\end{document}